\documentclass[12pt,a4paper,fleqn]{article}
\usepackage[DIV12]{typearea}
\usepackage{amsmath,amsfonts,amssymb}
\usepackage{graphicx}
\usepackage{cite}
\usepackage{mathrsfs}
\usepackage{bbm}
\usepackage{pifont}
\usepackage{units}
\usepackage[small,loose]{subfigure}  

\newcommand{\eVdist}{\kern-0.06em}

\DeclareMathOperator{\diag}{diag}

\DeclareMathOperator{\rank}{rank}

\newcommand{\CenterEps}[2][1]{\ensuremath{\vcenter{\hbox{\includegraphics[scale=#1]{#2.eps}}}}} 

\newcommand{\I}{\mathrm{i}}

\newcommand{\SO}[1]{\ensuremath{\mathrm{SO}(#1)}}
\newcommand{\SU}[1]{\ensuremath{\mathrm{SU}(#1)}}
\newcommand{\U}[1]{\ensuremath{\mathrm{U}(#1)}}
\newcommand{\Z}[1]{\ensuremath{\mathbbm{Z}_{#1}}} 

\allowdisplaybreaks[1]

\title{Patterns of remnant discrete symmetries}

\begin{document}

\begin{titlepage}

\begin{flushright}
TUM-HEP 730/09
\end{flushright}

\vspace*{1.0cm}

\begin{center}
{\Large\bf 
Patterns of remnant discrete symmetries
}

\vspace{1cm}

\textbf{
Bj{\"o}rn Petersen\footnote[1]{Email: \texttt{petersen@ph.tum.de}},
Michael Ratz\footnote[2]{Email: \texttt{mratz@ph.tum.de}},
Roland Schieren\footnote[3]{Email: \texttt{Roland.Schieren@ph.tum.de}}
}
\\[5mm]
\textit{\small
Physik-Department T30, Technische Universit\"at M\"unchen, \\
James-Franck-Stra\ss e, 85748 Garching, Germany
}
\end{center}

\vspace{1cm}

\begin{abstract}
We analyze patterns of remnant discrete symmetries that arise from $\U1^N$
theories by spontaneous breaking. We describe a simple, geometrical way to
understand these patterns and provide methods for identifying the discrete
symmetries and bringing them to the simplest possible form. Applications in GUT
and string model building are briefly discussed.
\end{abstract}

\end{titlepage}

\newpage

\section{Introduction}

Symmetries play a key role in our understanding of fundamental physics. While
forces originate from continuous symmetries, discrete symmetries turn out to
explain many important properties of matter. The perhaps most fundamental
examples are the reflection and conjugation symmetries $P$, $T$ and $C$, yet
there are other crucial $\Z{N}$ symmetries. Examples for such symmetries include
matter or $R$ parity in the minimal supersymmetric extension of the standard
model (MSSM) \cite{Farrar:1978xj}. Matter parity (or family reflection
symmetry~\cite{Dimopoulos:1981dw}) leads to a suppression of the proton decay
rate and explains the stability of the MSSM dark matter candidate, the lightest
supersymmetric particle (LSP). It can arise as a discrete \Z2 subgroup of a
baryon-minus-lepton-number symmetry $\U1_{B-L}$. One can break $\U1_{B-L}$ to
matter parity by giving vacuum expectation values (VEVs) to fields with $B-L$
charge $\pm2$. In \SO{10} grand unified theories (GUTs) one may switch on
appropriate components of a $\overline{\boldsymbol{126}}$-plet (see e.g.\
\cite{Mohapatra:1998rq}); in string theory an analogous breaking pattern can be
achieved~\cite{Lebedev:2007hv}.  Discrete symmetries might also be an important
ingredient for solving the flavor puzzle.

Obtaining discrete symmetries as remnants of gauge symmetries is, at a
fundamental level, motivated by anomaly considerations. While discrete
symmetries often are imposed ad hoc as global symmetries, it has been argued that
global symmetries are violated by quantum effects unless they originate
from a gauge symmetry via spontaneous symmetry breaking
\cite{Krauss:1988zc,Banks:1989ag,Preskill:1990bm}.

Here we focus on Abelian discrete symmetries i.e.\ \Z{N} groups. (For a recent discussion of how to obtain non-Abelian discrete symmetries by spontaneous breaking see 
\cite{Adulpravitchai:2009kd}.) It is well known
how to obtain a single \Z{N}-symmetry from a \U1; this will be reviewed below.
On the other hand, the most promising candidate for a consistent theory of
quantum gravity, string theory, typically provides us with models which
exhibit a large rank gauge symmetry. This symmetry has to be (spontaneously)
broken to the standard model in realistic vacua, which will generically lead to
a non-trivial set of remnant discrete symmetries. The purpose of this study is to
work out how such symmetries can be identified and put into a simple, i.e.\
canonical, form.

\section{Multiple $\boldsymbol{\Z{N}}$ symmetries from several U(1) factors}
\label{sec:obtain_Zs}

In this section we show how to determine the remnant symmetries if we break
multiple \U1 factors by a set of VEVs. Throughout this study we shall assume the
\U1 charges of all fields to be integers, which implicitly fixes our conventions
for charge normalization. We start with the easiest example of one \U1 and two
fields.

\subsection{Review: $\boldsymbol{\U1\to\boldsymbol{\Z{q}}}$}

Consider a theory with gauge group \U1 and two complex scalar fields
$\phi$ and $\psi$ (cf.\ \cite{Krauss:1988zc}). Under the \U1 the fields
transform according to
\begin{subequations}\label{eq:easy_trafo_psi}
\begin{eqnarray} 
 \phi &\rightarrow& \mathrm{e}^{\I\, q\, \alpha(x)}\, \phi\;, \\
 \psi &\rightarrow& \mathrm{e}^{-\I\, \alpha(x)}\, \psi 
\end{eqnarray}
\end{subequations}
with $q \in \mathbbm{N}$, i.e.\ $\phi$ has charge $q$ and $\psi$ has charge $-1$.
The terms 
\begin{equation}
 \phi^*\phi\;,\quad\psi^*\psi\;,\quad
 \phi^*\phi\,\psi^*\psi \quad \text{and} \quad  \phi\, \psi^q + \text{h.c.} \label{eq:interactions}
\end{equation}
and powers as well as products thereof are gauge invariant. Suppose now that
$\phi$ acquires a VEV. This leaves us with effective interaction terms of the
form $(\psi^q)^n$ with $n\in\mathbbm{N}$, dictated by the symmetry
\begin{equation}\label{eq:remaing_baby_symmetry}
 \psi~\rightarrow~\mathrm{e}^{2\pi\I\, \ell/q}\, \psi 
 \qquad \text{with} \quad 
 \ell=0,1,\ldots q-1\;. 
\end{equation}
An equivalent way to obtain this result is stating that the remaining symmetry is
determined by the condition
\begin{subequations}\label{eq:baby_breaking_condition}
\begin{equation} 
 \mathrm{e}^{\I\, q \,\alpha(x)}\,\phi~=~
 \phi \quad \Rightarrow \quad 
  q\, \alpha~=~ 2\pi \, \,\ell\quad \text{with} \quad  
 \ell \in\mathbbm{Z}\;.
\end{equation}
Hence, 
\begin{equation}\label{eq:baby_breaking_condition2}
 \frac{q\,\alpha}{2\pi}~=~0\mod1\;,
\end{equation}
\end{subequations}
or, equivalently, $\alpha\in\frac{2\pi}{q}\mathbbm{Z}$.
We have just rederived the well known result that, by giving a VEV to a field
with charge $q$, the \U1 gets broken to a \Z{q} discrete subgroup. In what follows
we will generalize this to situations in which several \U1s get broken to a
number of \Z{n}s.

\subsection{The general case}

Let us now consider the general case of a $\U1^N$ gauge theory with $M$ scalar
fields $\phi^{(i)}$ $(1\le i\le M)$, which will acquire VEVs, and $K$ other
`matter' fields $\psi^{(j)}$ ($1\le j\le K)$. We will denote the charge of the
fields w.r.t.\ the $j^\mathrm{th}$ \U1 factor by $q_j(\phi^{(i)})$ and
$q_j(\psi^{(i)})$ respectively. Accordingly, the $\phi^{(i)}$ fields transform as
\begin{equation}
 \phi^{(i)} 
 ~\rightarrow~\exp\left(\I\,\sum_j q_j(\phi^{(i)})\, \alpha_j(x)\right)\, 
 \phi^{(i)}\;.
\end{equation}
$q(\phi^{(i)})$ can be thought of as an $N$-dimensional charge vector and
$Q_{ij}=q_j(\phi^{(i)})$ as an $M\times N$ charge matrix.  Suppose now that
$N>\rank Q$. In this case, there are unbroken \U1 factors. Then we can rotate
the \U1 directions by an orthogonal transformation such that all $\phi^{(i)}$
will be uncharged under $(N-\rank Q)$ \U1 factors. These \U1 factors will not be
affected by the VEVs of the $\phi^{(i)}$ fields and we do not have to consider them
any further.  Therefore, without loss of generality, we will from now on
consider the case $N\leq\rank Q$. 
Notice also that in supersymmetric theories the rank of the charge matrix 
cannot be maximal as $D$-flatness requires a non-trivial solution of $\sum_i
n_i\,q(\phi^{(i)})~=~0$ with $n_i\in\mathbbm{N}_0$.

To identify the remnant discrete symmetries after spontaneous symmetry breaking,
consider the generalization of equation \eqref{eq:baby_breaking_condition} in
our simple example,
\begin{subequations}
\begin{equation}
 \exp\left( \I\,\sum_j  q_j(\phi^{(i)})\, \alpha_j\right)\, \phi^{(i)}
 ~\stackrel{!}{=}~
 \phi^{(i)}\;.
\end{equation}
This is equivalent to 
\begin{equation} \label{eq:ConstraintsOnAlpha}
 \sum_j q_j(\phi^{(i)})\, \alpha_j
 ~\stackrel{!}{=}~2\pi \, \,\ell^i \quad \text{with}
 \quad  \ell^i \in\mathbbm{Z}\;.
\end{equation}
As in \eqref{eq:baby_breaking_condition2}, the right-hand side represents the
usual `mod conditions' for discrete breaking.
Equation~\eqref{eq:ConstraintsOnAlpha} can be recast in matrix notation,
\begin{equation}\label{eq:breaking_condition}
 Q\,\alpha~\stackrel{!}{=}~2\pi\,\ell
 \quad\text{with}\quad\ell\in\mathbbm{Z}^M
 \;. 
\end{equation}
\end{subequations}
Recall that $Q$ is an $M\times N$ matrix with elements in $\mathbbm{Z}$. Such a
matrix can always be brought into diagonal form by two unimodular
transformations, i.e.\ invertible matrices over $\mathbbm{Z}$. Concretely, there
exist $A\in\text{GL}(M,\mathbbm{Z})$ and $B\in\text{GL}(N,\mathbbm{Z})$ such
that
\begin{equation}\label{eq:SmithNormalForm}
 A\,Q\,B
 ~=~
 D
 ~=~\diag'(d_1,\ldots,d_N)  
 \qquad \text{and} \qquad d_{i} ~\text{divides}~ d_{i+1} \;.
\end{equation}
Here $\diag'$ means that $D$ is an $M\times N$ matrix whose only non-zero
elements $d_i$ are at the $ii$ positions. There are $N$ non-zero diagonal entries due to $\text{rank}Q \ge N$.  
$D$ is called Smith normal form (or just `normal form' as in \cite{Jacobsen}) of
$Q$. Note that unimodular matrices have two important properties:
\renewcommand{\labelenumi}{\arabic{enumi}.}
\begin{enumerate}
 \item They have determinant $\pm 1$.\footnote{If  the determinant was zero, the
  matrices would not be invertible. If the absolute value of the determinant was
  greater than 1, the inverse would not be an integer matrix, i.e.\ the matrix
  would not be invertible over $\mathbbm{Z}$.} 
  \label{SLN1}
 \item The greatest common divisor of all matrix elements in a single row is 1
  because otherwise the determinant would not be $\pm 1$. The same applies to
  each column.
  \label{SLN2}
\end{enumerate}
\renewcommand{\labelenumi}{\arabic{enumi}.}
Transformation \eqref{eq:SmithNormalForm} allows us to rewrite
\eqref{eq:breaking_condition},
\begin{equation}
 A^{-1}\,D\,B^{-1}\,\alpha~=~2\pi\,\ell\;.
\end{equation}
Now multiply this equation by $A$. Due to the second property of unimodular
matrices, the `mod conditions' in equation~\eqref{eq:breaking_condition} remain
unchanged, since $\ell'=A\,\ell$ still takes all values in $\mathbbm{Z}^M$ if
$\ell$ does. Defining $\alpha'=B^{-1}\,\alpha$ we arrive at
\begin{equation} \label{ellprime}
 \alpha'_j~=~2\pi \frac{\ell'^j}{d_j} 
 \qquad \text{with} \qquad  
 0\le\ell'^j\le d_j-1 \;.
\end{equation}
Hence, we see that the remnant discrete symmetry is $\Z{d_1}\times \ldots \times
\Z{d_N}$. If there are some $d_i=1$, the corresponding factors are trivial and
can be omitted. The fields $\psi^{(j)}$ then transform according to
\begin{equation}
 \psi^{(j)}
 ~\rightarrow~
 \exp\left( \I\, \sum_{k,\ell} q_k(\psi^{(j)})\, B_{k\ell}\, \alpha'_\ell
 \right) \psi^{(j)}
 ~=~ 
 \exp\left( 2\pi\I\, \sum_{k} q'_k(\psi^{(j)})\,
 \frac{\ell'^k}{d_k} \right)\, \psi^{(j)}
\end{equation}
with new charges $q'_k(\psi^{(j)})=\sum_i q_i(\psi^{(j)})\, B_{ik}$, which are
defined modulo $d_k$. That is, we can choose
$q'_k(\psi^{(j)})\in\{0,\ldots,d_k-1\}$.

\subsection{An example with two U(1) factors}
\label{sec:another_example}

Let us illustrate the above procedure by an example. Consider a  $\U1\times\U1'$
theory with three fields obtaining VEVs and two other fields.  That is, we have
$N=2$, $M=3$ and $K=2$. The charges are given in table \ref{tab:charges}.
\begin{table}[h]
\centerline{
\subtable[VEV fields.\label{tab:Example1VEVs}]{
\begin{tabular}{ccc}
& $\U1$ &  $\U1'$\\
$\phi^{(1)}$ & 8 & -2 \\
$\phi^{(2)}$ & 4 & 2 \\
$\phi^{(3)}$ & 2 & 4 
\end{tabular}
}\quad
\subtable[Matter fields.]{
\begin{tabular}{ccc}
& $\U1$ &   $\U1'$ \\
 $\psi^{(1)}$ & 1 & 3 \\ 
 $\psi^{(2)}$ & 1 & 5 \\
\end{tabular}
}
}
\caption{Charges of the fields with respect to the two \U1 factors.}
\label{tab:charges}
\end{table}
In this example, we only consider scalar fields, such that we do not have to
worry about anomalies. Later, in the applications in
section~\ref{sec:applications} we will discuss supersymmetric, anomaly-free
settings.
The charge matrix is given by the charges of the VEV fields (cf.\
table~\ref{tab:charges}~(a)),
\begin{equation}
 Q~=~\left( 
 \begin{array}{cc}
  8 & -2 \\ 
  4 & 2 \\
  2 & 4
 \end{array}
 \right) \;.
\end{equation}
The diagonal matrix $D$ and the transformation matrix $B$ are
\begin{equation}\label{eq:diagonalisation}
 D~=~\left( 
 \begin{array}{cc}
 2 & 0 \\ 
 0 & 6 \\
 0 & 0
 \end{array}
 \right) 
 \qquad
 \text{and}
 \qquad
 B~=~\left( 
 \begin{array}{cc}
 1 & -2 \\ 
 0 & 1 
 \end{array}
 \right)\;.
\end{equation}
Hence, we can read off that we are left with a $\Z{2}\times\Z{6}$ symmetry. The charges of
the $\psi^{(i)}$ fields can be determined by multiplying their charge matrix by $B$
from the right,
\begin{equation}
 \left( 
 \begin{array}{cc}
 1 & 3 \\ 
 1 & 5 
 \end{array}
 \right)\, 
 \left( 
 \begin{array}{cc}
 1 & -2 \\ 
 0 & 1
 \end{array}
 \right)
 ~=~
 \left( 
 \begin{array}{cc}
 1 & 1 \\ 
 1 & 3 
 \end{array}
 \right)\;.
\end{equation}
The new charges of the $\psi^{(i)}$ fields are given by the rows of this matrix.
Altogether we find that the setting discussed here leads to a $\Z{2}\times\Z{6}$
symmetry, under which $\psi^{(1)}$ has charge (1,1) and  $\psi^{(2)}$ has charge
(1,3).

\subsection{Visualization}
\label{sec:visualisation}

We will now provide a simple, geometrical way of envisaging the symmetry
breaking patterns. First, notice that a direct product of groups
$\Z{d_1}\times\ldots\times\Z{d_N}$ can be represented by an $N$-dimensional
lattice. Each element of the group can be thought of as a point in the
fundamental region or unit cell of the lattice. The volume of the unit cell is
the number of elements and the order of the group. 

\begin{figure}[h]
\centerline{
\subfigure[Original lattice.\label{fig:ChargeLattice1a}]{\CenterEps{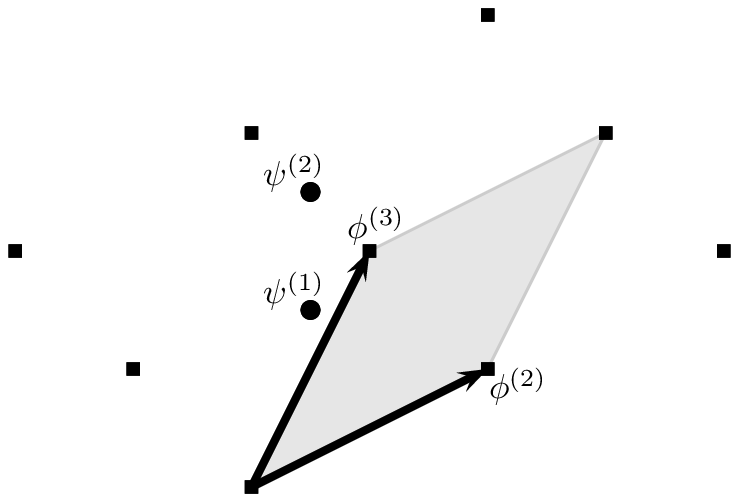}}
\qquad\qquad
\subfigure[Diagonalized lattice.\label{fig:ChargeLattice1b}]{\CenterEps{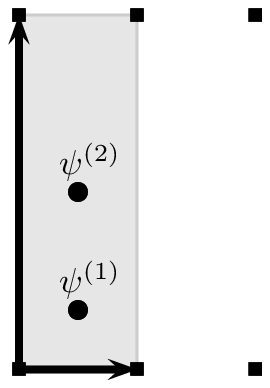}}
}
\caption{Illustration of discrete breaking.}
\label{fig:ChargeLattice1}
\end{figure}

Let us illustrate this by the above example. The VEV fields $\phi^{(i)}$ with
charges as listed in table~\ref{tab:charges}~(a) span the lattice. Since the
first VEV field is an integer linear combination of the latter two,
$\phi^{(1)}=3\phi^{(2)}-2\phi^{(3)}$, $\phi^{(2)}$ and $\phi^{(3)}$ span a basis
of the charge lattice as illustrated by the arrows in
figure~\ref{fig:ChargeLattice1a}. The matter fields $\psi^{(i)}$ are represented
by the bullets. A coupling $(\psi^{(1)})^{n_1}\,(\psi^{(2)})^{n_2}$ is allowed by
the discrete symmetries if and only if $n_1\,q(\psi^{(1)})+n_2\,q(\psi^{(2)})$
lies on a node in the charge lattice, which are represented by squares. 

Our procedure described in section~\ref{sec:another_example} amounts to finding
an orthogonal basis for the lattice which is given by the rows of the diagonal
matrix $D$. The matrix $A$ performs a rotation of the $\phi^{(i)}$ charges which eliminates linear dependencies, while $B$ is the transformation between the bases. In the new
basis, the lattice is orthogonal, and the $\psi^{(i)}$ charges are given by the
projections on the basis vectors.  We see that there is a $\Z2\times\Z6$
discrete symmetry where the \Z2 corresponds to the horizontal and the \Z6 to the
vertical direction in figure~\ref{fig:ChargeLattice1b}. The matter fields have
charges $(1,1)$ and $(1,3)$, respectively, which are of course nothing but the
coordinates of the bullets in figure~\ref{fig:ChargeLattice1b}. Note, however,
that the true symmetry is smaller than that; we will describe in
section~\ref{sec:Simplification} how to eliminate potential redundancies.

\subsection{Inverting the problem}
\label{sec:Inversion}

Often one is interested in `inverting' the above procedure. Instead of
determining a discrete symmetry that is left after certain fields attain VEVs,
one would like to understand which fields need to be switched on in order to
obtain a desirable discrete symmetry. Examples for such symmetries include
matter parity and proton hexality~\cite{Dreiner:2005rd}. We describe now a
simple algorithm for accomplishing this task. We focus on the case of a remnant
\Z2 symmetry; the extension to the general case is straightforward.
\begin{dingautolist}{'300}
 \item Start with a set `matter' fields $\psi^{(i)}$  with charge vectors
 $q(\psi^{(i)})$.
 \item The $q(\psi^{(i)})$ span a lattice 
  \begin{equation}
   \Lambda~=~\left\{ \sum_\alpha n_\alpha\,\lambda^{(\alpha)};~n_\alpha\in\mathbbm{Z}\right\}
  \end{equation}
  such that $q(\psi^{(i)})\in\Lambda$, i.e.\
  \[
   q(\psi^{(i)})~=~\sum_\alpha n^{(i)}_\alpha\,\lambda^{(\alpha)}
   \quad\text{with}~n^{(i)}_\alpha\in\mathbbm{Z}
  \]
  and there is no coarser lattice with the same property.\footnote{This
  lattice can, for instance, be obtained with the \texttt{Mathematica} command
  `\texttt{LatticeReduce}'.}
 \item Compute the dual lattice 
  \begin{equation}
   \Lambda~=~\left\{ \sum_\beta n^\beta\,\lambda_{(\beta)}^*;~n^\beta\in\mathbbm{Z}\right\}
  \end{equation}
  with $\lambda_{(\beta)}^*\cdot\lambda^{(\alpha)}=\delta^\alpha_\beta$. The
  basis vectors $\lambda_{(\beta)}^*$ have the obvious property that $\lambda_{(\beta)}^*\cdot
  q(\psi^{(i)})=n^{(i)}_\beta\in\mathbbm{Z}$.
 \item Now try to build linear combinations 
  \[ \mathsf{t}_\Delta~:=~\sum_\beta m^\beta\,\lambda_{(\beta)}^*\] 
  such that
  \begin{equation}\label{eq:tDeltaProperty}
   \mathsf{t}_\Delta\cdot q(\psi^{(i)})~=~1\mod 2\;.
  \end{equation}
  It is obvious that one just has to scan all possible combinations with
  $m^\beta\in\{0,1\}$ since an even $m^\beta$ will always lead to an even number
  on the right-hand side of \eqref{eq:tDeltaProperty}. $\mathsf{t}_\Delta$ is
  then unique up to \U1 generators under which all $\psi^{(i)}$ are neutral.
\item Given a generator $\mathsf{t}_\Delta$, one has to check whether the model
 contains fields $\phi^{(i)}$ with charges of the type `even over odd'.
 If this is the case, switching on the $\phi^{(i)}$ fields yields configurations
 with a \Z2 symmetry under which the $\psi^{(i)}$ fields are odd.
\end{dingautolist}
Only the last two steps have to be slightly modified in order to obtain an arbitrary
\Z{N} symmetry. In section~\ref{sec:StringModelBuilding} we will apply these
methods in order to identify phenomenologically attractive string vacua.

\section[Simplifying multiple $\Z{N}$ symmetries]{Simplifying multiple
$\boldsymbol{\Z{N}}$ symmetries}
\label{sec:Simplification}

In this section we will consider a finite Abelian group
$G=\Z{d_1}\times\ldots\times\Z{d_N}$ with $K$ fields $\psi^{(i)}$,
$i=1,\ldots,K$, transforming under $G$. This setup may or may not be a result of
the diagonalization procedure described in section
\ref{sec:obtain_Zs}.\footnote{That is, we do not require $d_i | d_{i+1}$ here.}
Our aim is to eliminate redundancies, i.e.\ make the discrete symmetry `as
simple as possible'.

First, consider a toy example which consists of a \Z{6}-symmetry with one field
$\psi$ with charge 4. This is equivalent to a \Z{3}-symmetry where $\psi$ has
charge 2. In the case of one \Z{N}-symmetry with fields $\psi^{(i)}$ and charges
$q(\psi^{(i)})$, it is easy to see if there is some redundancy in our
description of the symmetry. If the greatest common divisor (GCD) of the
$q(\psi^{(i)})$ and $N$ is greater than one, we can divide $q(\psi^{(i)})$ and
$N$ by their GCD. That is precisely what happened in our above toy example. This
idea of reducing the group order carries over to the general case. In addition,
the method we are going to develop will bring $G$ into a canonical form. Every
finite Abelian group can be written as a direct product of the form 
\begin{equation}
 G_\mathrm{canonical}
 ~=~
\Z{d_1}\times\ldots\times\Z{d_N}\;,
\qquad \text{where}~d_{i}~\text{divides}~d_{i+1}\;.
\end{equation}
The $d_i$ are uniquely determined by the group and are called invariant factors.
The diagonalization process described in section \ref{sec:obtain_Zs} always
leads to this form (cf.\ equation~\eqref{eq:SmithNormalForm}). 

\subsection{The general case}

Let us look at the charge matrix $(Q_{\psi})_{ij}=q_j(\psi^{(i)})$.   This is a
$K \times N$ matrix with integer entries. Again, the idea is to calculate the
Smith normal form of $Q_{\psi}$. However, to get a meaningful result, we need to bring
$G$ into the form $\Z{d}\times\ldots\times\Z{d}$ where $d$ is the least common
multiple (LCM) of the $d_i$. This is because a $\Z{d}^N$ discrete symmetry
allows us to perform discrete rotations of the generators. Enlarging the
symmetry implies a rescaling of the charge matrix,
$(Q_{\psi}')_{ij}=\frac{d}{d_j}\,q_j(\psi^{(i)})$. The fields transform now
according to
\begin{equation}
 \psi^{(i)} 
 ~\rightarrow~ 
 \exp\left(2\pi\I\,\sum_j  (Q_{\psi}')_{ij}\, \alpha_j\right)\, \psi^{(i)} 
 \qquad \text{where} \quad 
 \alpha_j~=~\frac{\ell_j}{d} \quad \text{with} \quad 
 0\le\ell_j\le d-1\;.
\end{equation}
$(Q_{\psi}')$ can be brought into Smith normal form $S$ by unimodular transformations 
$E\in\text{GL}(K,\mathbbm{Z})$ and $F\in\text{GL}(N,\mathbbm{Z})$,
\begin{equation}
 E\,Q_{\psi}'\,F~=~S~=~\diag'(s_1,\ldots,s_k) 
 \qquad \text{where} \quad k~=~\min(K,N)\;.
\end{equation}
If $\rank Q_\psi'<k$, some $s_j$ might vanish.
The corresponding rows do not have to be considered further.
This yields the transformation behavior
\begin{eqnarray}
 \psi^{(i)}
 & \rightarrow &
 \exp\left(2\pi\I\,\sum_{j,m,n}  E^{-1}_{ij}\, S_{jm}\,
 F^{-1}_{mn}\,\alpha_n\right)\, \psi^{(i)} 
 \nonumber\\
 & = & \exp\left(2\pi\I\,\sum_{j,n} E^{-1}_{ij}\, s_j\, F^{-1}_{jn}\,
 \frac{\ell_n}{d}\right)\, \psi^{(i)} \;.
\end{eqnarray}
Just like in equation (\ref{ellprime}), we are allowed to define $\ell'_j=\sum_n F^{-1}_{jn}\, \ell_n$. Hence, we get
\begin{equation}
 \psi^{(i)}~\rightarrow~ \exp\left(2\pi\I\,\sum_{j}  E^{-1}_{ij}\,
  \frac{\ell'_j}{d/s_j}\right)\, \psi^{(i)} \;.
\end{equation}
This tells us that we have rewritten our symmetry group $G$ as
$\Z{d'_1}\times\ldots\times\Z{d'_k}$ where $d'_i$ is the numerator of the
(reduced) fraction $\frac{d}{s_i}$ and $d'_{i+1}$ divides $d'_i$. If a $d'_i$ is
equal to one, this factor can be omitted. The new charges of the fields,
$q''(\psi^{(i)})$, are encoded in the matrix $E^{-1}$, 
\begin{equation}
 E^{-1}~=~\left(\begin{array}{c}
  q''(\psi^{(1)})\\
  q''(\psi^{(2)})\\
  \vdots
 \end{array}\right)
 \;. 
\end{equation}
These charges are equivalent to the the discrete charges $s_i'\,q'(\psi^{(i)})$
(with $s_i'$ denoting the denominator of the reduced fraction $d/s_i$),
which one immediately reads off. An obvious consequence of our discussion is
that after simplification there are at most as many \Z{d_i} factors as fields. 
Note that the volumes of the fundamental region of $G$'s lattice ($V_G = \prod^{N}_{i=1} d_i$) and the lattice
spanned by the $\psi^{(i)}$ charges ($V_\psi = \text{det}Q_\psi$) provide a necessary but not sufficient
criterion for redundancies: if both possess a GCD, the size of the symmetry may
be reduced by this GCD.

\subsection{An alternative derivation}

Before discussing an example, let us present an alternative point of view. A
coupling $(\psi^{(1)})^{n_1}\cdots(\psi^{(K)})^{n_K}$ is allowed by the discrete
symmetry $G=\Z{d_1}\times\ldots\times\Z{d_N}$ only if there is a vector
$n\in\mathbbm{Z}^K$ such that
\begin{equation} \label{eq:AllowedCoupling}
 Q_\psi^T\,n~=~\diag(d_1,\dots,d_N)\,m
\end{equation}
with some $m\in\mathbbm{Z}^N$ and $Q_\psi^T=\bigl(q(\psi^{(1)}),\dots , 
q(\psi^{(K)})\bigr)$. Equation~\eqref{eq:AllowedCoupling} can be rewritten as
\begin{equation}
 \diag\left(\frac{d}{d_1},\dots,\frac{d}{d_N}\right)\cdot Q_\psi^T\,n
 ~=~d\,m\;,
\end{equation}
where $d$ denotes the LCM of the $d_i$, as before. Now we diagonalize the matrix
on the left hand side of the equation,
\begin{equation}
 \diag\left(\frac{d}{d_1},\dots,\frac{d}{d_N}\right)\cdot Q_\psi^T
 ~=~(F^{-1})^T\cdot S\cdot (E^{-1})^T
\end{equation}
with the $N\times K$ matrix $S=\diag'(s_1,\dots s_k)$, the unimodular matrices
$E$ and $F$, and $k=\min(K,N)$, as before.   Let $\nu$ denote the rank of
$Q_\psi$, i.e.\ $S=\diag'(s_1,\dots s_\nu,0,\dots)$. Now
\eqref{eq:AllowedCoupling} can be recast as
\begin{equation}
 \diag'\left(\frac{s_1'}{d_1'},\dots,\frac{s_\nu'}{d_\nu'},0,\dots\right)
 \,(E^{-1})^T\,n
 ~=~
 m'\;,
\end{equation}
where $m'=F^T\,m$, and again $s_i'$ and $d_i'$ denote the
numerators and denominators of the reduced fractions $s_i/d$, respectively.
The rank of the matrix on the left-hand side of this equation is $\nu$.
We can therefore truncate the equation, 
\begin{equation}
 \diag\left(\frac{s_1'}{d_1'},\dots,\frac{s_\nu'}{d_\nu'}\right)
 \,(E_\nu^{-1})^T\,n
 ~=~
 m_\nu\;,
\end{equation}
where $(E_\nu^{-1})$ denotes the left $\nu$ columns of $E^{-1}$ (such that
$(E_\nu^{-1})^T$ is the upper $\nu\times K$ part of $(E^{-1})^T$),
and $m_\nu\in\mathbbm{Z}^\nu$. This equation is equivalent to
\begin{equation}
 (E_\nu^{-1})^T\,n
 ~=~
 \diag'\left(\frac{d_1'}{s_1'},\dots,\frac{d_\nu'}{s_\nu'}\right)\,m_\nu\;,
\end{equation}
Comparing this equation with  \eqref{eq:AllowedCoupling} reveals that the rows
of $E_\nu^{-1}$ contain the charges of the $\psi^{(i)}$ and the $d_i'$ determine the
canonical symmetry. 

\subsection{Example}

Let us continue the example of section \ref{sec:another_example}. After breaking
both \U1s we are left with a $\Z{2}\times\Z{6}$ and a charge assignment given in
table \ref{tab:example}.
\begin{table}[h]
\centerline{\subtable[Original charges.]{
\begin{tabular}{ccc}
& \Z{2} &  \Z{6}   \\ 
$\psi^{(1)}$ & 1 & 1 \\ 
$\psi^{(2)}$ & 1 & 3  
\end{tabular}
}
\quad
\subtable[`Blown up' charges.]{
\begin{tabular}{ccc}
& \Z{6} &  \Z{6}   \\ 
$\psi^{(1)}$ & 3 & 1 \\ 
$\psi^{(2)}$ & 3 & 3  
\end{tabular}
}
\quad
\subtable[Minimal charges.]{\quad
\begin{tabular}{cc}
& \Z{6}    \\ 
$\psi^{(1)}$  & 1 \\ 
$\psi^{(2)}$  & 3  
\end{tabular}\quad
}
}
\caption{The example from section \ref{sec:another_example} continued.
The original charges (a) are blown up to the charges of an extended
$\Z6\times\Z6$ symmetry, which can be reduced to a \Z6 symmetry by discrete
rotations.}
\label{tab:example}
\end{table}
We have to extend  our symmetry to $\Z{6}\times\Z{6}$. The charge matrix and the
Smith normal form are
\begin{equation}
 Q_{\psi}'~=~
\left( 
\begin{array}{cc}
3 & 1 \\ 
3 & 3 
 \end{array}
\right) 
~=~
\left( 
\begin{array}{cc}
1 & 0 \\ 
3 & -1 
 \end{array}
\right)\,
\left( 
\begin{array}{cc}
1 & 0 \\ 
0 & 6 
 \end{array}
\right)\,
\left( 
\begin{array}{cc}
3 & 1 \\ 
1 & 0 
 \end{array}
\right)
~=~E^{-1}\,S\,F^{-1}
\;.
\end{equation}
The $d'_i$ can be inferred from the diagonal matrix $S$: they are given by 6
times the inverses of the diagonal entries, i.e.\ we have $d'_1=6$ and
$d'_2=1$. The numerators are hence 6 and 1, such that we are left
with a $\Z{6}\times\Z1=\Z{6}$ symmetry. The charges are given by the rows of
$E^{-1}$ (modulo 6); since one factor is trivial we obtain that $\psi^{(1)}$
has charge 1  and $\psi^{(2)}$ has charge 3.

\subsection{Visualization}

\begin{figure}[htp]
\centerline{\subfigure[Extended lattice.]{\CenterEps{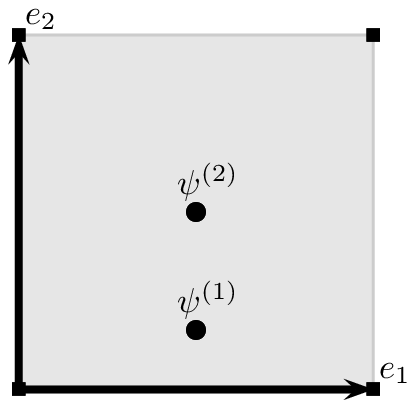}}\quad
\subfigure[Simplified lattice.]{\CenterEps{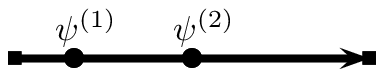}}}
 \caption{(a) Extended and (b) simplified charge lattices. 
 }
 \label{fig:lattice3}
\end{figure}

The visualization works as before. We extend the lattice to $\Z6\times\Z6$
(figure~\ref{fig:lattice3}~(a)). 
Our simplification process amounts to identifying a direction
on which both nodes lie (figure~\ref{fig:lattice3}~(b)).
It is easy to see that $\psi^{(1)}$ sits at 1/6 of the length of the
one-dimensional $\Z6$ lattice, corresponding to $\Z6$ charge 1, while
$\psi^{(2)}$ sits at $1/2=3/6$ of the length, which leads to $\Z6$ charge 3.

\section{Automatization}

We provide a \texttt{Mathematica} package which automatically identifies the
remnant symmetries, as discussed in section~\ref{sec:obtain_Zs}, and brings them to the
canonical form, as described in section~\ref{sec:Simplification}, on our web
site~\cite{ZNcode:2009zn}. This package has been used in our applications, which
will be discussed in what follows.

\section{Applications}
\label{sec:applications}

\subsection{GUT model building}
\label{sec:GUTmodelBuilding}

Let us now apply the above methods to model building. We focus on a specific GUT
model, which has been discussed in \cite{Mohapatra:2007vd}. There, an anomaly-free
\Z{6} symmetry was found that may allow us to suppress proton decay in \SO{10}
GUTs. The field content is given in table \ref{tab:SO10_field_content}.
\begin{table}[htp]
 \centering
\begin{tabular}{cccccccc}
        & $\psi_m$ & $H$ & $H'$ & $\psi_H$ & $\overline{\psi}_H$ & $A$ & $S$   \\ 
\SO{10} & \textbf{16}& \textbf{10} & \textbf{10} & \textbf{16} & $\overline{\textbf{16}}$ & \textbf{45} & \textbf{54} \\ 
\Z{6}   & 1 & -2 & 2 & -2 & 2 & 0 & 0
\end{tabular}
 \caption{Field content of an \SO{10} model. }
 \label{tab:SO10_field_content}
\end{table}
There are three generations $\psi_m$ of standard model matter; $H$ and $H'$
contain the standard model Higgses. The other fields are used to break \SO{10}
down to the standard model gauge group $G_\mathrm{SM}$.  Since \SO{10} has rank
5 there is a \U1 factor, called $\U{1}_\chi$, in addition to the standard model
gauge group. This $\U{1}_\chi$ gets broken by giving VEVs to the SM singlet
fields contained in $\psi_H$ and $\overline{\psi}_H$, which have $\U1_\chi$
charges $\pm5$. Hence, we have a situation where a $\U{1}_\chi\times\Z{6}$
symmetry gets broken by a field with charges $(\pm5,\pm2)$. 
We can apply the routine described in section \ref{sec:obtain_Zs}. This can be
accomplished by extending the \Z6 to a \U1, introducing a dummy field with
charges (0,6) and assigning a VEV to this field.

We find that the $\U{1}_\chi\times\Z{6}$ gets broken to a \Z{30} with charges
given in table \ref{tab:Z30_charges}.
\begin{table}[htp]
 \centering
\begin{tabular}{cccccccc}
        & $Q$ & $\Bar{U}$ & $\Bar{D}$ & $L$ & $\Bar{E}$ & $H_u$ & $H_d$   \\ 
\Z{30} & 3& 3 & 11 & 11 & 3 & 24 & 16 \\
\Z{5}$\times$\Z{6} & (1,1) & (1,1) & (2,5) & (2,5) & (1,1) & (3,4) & (2,0) 
\end{tabular}
 \caption{The $\Z{30} \simeq \Z{5}\times\Z{6}$ charges of the MSSM field content.}
 \label{tab:Z30_charges}
\end{table}
Let us remark that the \Z5 subgroup of this \Z{30} is redundant in the following
sense: whenever a coupling is gauge invariant under $G_\mathrm{SM}$, the
coupling is also invariant under \Z5. This is because \SU5 has a non-trivial
center \Z5 (cf.\ the analogous discussion in \cite{Csaki:1997aw}).  Note also
that $G_\mathrm{SM}$ invariance is the same as \SU5 invariance since the Cartan
generators are equivalent. Our results imply that, if we assume that all states
except the ones listed in table~\ref{tab:Z30_charges} attain masses, we are left
with an anomalous field content. Specifically, the three generations of SM
matter are anomaly free, but the \Z6 charges of the Higgs fields exhibit an
anomaly (for a discussion of discrete anomaly constraints
see~\cite{Ibanez:1991hv,Banks:1991xj,Araki:2008ek}). Using discrete anomaly matching
\cite{Csaki:1997aw} we can hence infer that the above assumption is inconsistent:
either further light states have to be present, or the \Z6 cannot be exact,
i.e.\ we have to introduce further fields that attain VEVs. Of course
approximate symmetry might be sufficient for suppressing the dangerous dimension
five operators, and might well be correlated with flavor hierarchies (cf.\ the
discussion in~\cite{Azatov:2008vu}). On the other hand, our findings show that
the  \Z{6} symmetry introduced in the \SO{10} GUT in \cite{Mohapatra:2007vd}
cannot give rise to proton hexality~\cite{Dreiner:2005rd}.

\subsection{String model building}
\label{sec:StringModelBuilding}

As mentioned in the introduction, an exact matter parity can also be obtained in
string theory by breaking a $\U1_{B-L}$ symmetry by two
units~\cite{Lebedev:2006kn,Lebedev:2007hv,Buchmuller:2008uq}. This has led to a
couple of vacuum configurations with an exact matter parity. Yet it turns out
that, with this strategy, one is not always successful: within the so-called
mini-landscape \cite{Lebedev:2006kn} of heterotic orbifolds with exact MSSM
spectra, vacuum configurations with an exact matter parity could only be
identified in a small fraction of the models; that is, an appropriate
$\U1_{B-L}$ symmetry could only be identified in 15 out of 218 possible models.
The obstacles encountered in this $\U1_{B-L}$-based approach are perhaps best
illustrated in a concrete example. The model discussed in
\cite{Buchmuller:2005jr} (which later became absorbed in the mini-landscape) does
have a $\U1_{B-L}$ symmetry~\cite{Buchmuller:2006ik,Buchmuller:2008uq}, yet in
the 4D zero mode spectrum there is no field with an even $B-L$ charge (nor with
a fractional charge of the type `even over odd', which is also sufficient to provide us
with matter parity~\cite{Lebedev:2007hv}).\footnote{In~\cite{Buchmuller:2008uq},
therefore, an alternative has been discussed where the $\U1_{B-L}$ gets broken
by fields that do not appear in the 4D zero-mode spectrum but by states that are
massless in an intermediate 6D orbifold GUT picture and get projected out by
going to 4D. This requires a cancellation between a Kaluza-Klein mass and some
non-trivial vacuum expectation values of certain standard model singlets.}

On the other hand, it is also clear that one does not really need a $\U1_{B-L}$
symmetry with the standard charges for the MSSM matter fields. Any \U1 symmetry
under which the matter states have odd charges, and for which there exist SM
singlets with even charges, could do the job. 
Using the methods discussed in section~\ref{sec:Inversion}, we were able to
identify a collection of $G_\mathrm{SM}$ invariant fields $\phi^{(i)}$ that
break the \U1 factors down to matter parity in the model presented in
\cite{Buchmuller:2005jr,Buchmuller:2006ik}. It is given by
\begin{eqnarray}
 \{\phi^{(i)}\}
 & = &
 \{s_{1},s_{2},s_{3},s_{5},s_{7},s_{9},s_{12},s_{14},s_{16},s_{18},s_{19},s_{20},s_{22},s_{23},s_{24},s_{34},s_{39},s_{40},
 \nonumber\\
 & & \phantom{\{}
 s_{41},s_{48}, s_{53},s_{54},s_{57},s_{58},s_{59},s_{60},s_{61},s_{62},s_{65},s_{66},
 f_{1-4},\bar f_{1-4}, h_{1-14}\}
\end{eqnarray}
in the notation of~\cite{Buchmuller:2006ik}. Our algorithm gives us a \Z{10}
symmetry, however, as discussed in section~\ref{sec:GUTmodelBuilding}, the \Z5,
which is just the non-trivial center of \SU5, is redundant. A quick scan
indicates that in many (if not in all) mini-landscape models vacua with matter
parity can be obtained. A detailed analysis of these issues and of the
phenomenological properties of such vacua will be carried out elsewhere.

\section{Discussion}

We have described a simple method to determine symmetry breaking patterns which
arise when $\U1^N$ gauge theories get broken to discrete subgroups. This
method has a very simple geometrical interpretation: the fields acquiring VEVs
define a charge lattice. Couplings $(\psi^{(1)})^{n_1}\,(\psi^{(2)})^{n_2}\dots$
are only allowed by the remnant discrete symmetries if the sum of the charge
vectors, $n_1q(\psi^{(1)})+n_2q(\psi^{(2)})+\dots$, lies on a node of the charge
lattice. 

Unimodular transformations allow us to identify the remnant discrete
symmetries, and  to make them as simple as possible, i.e.\ to determine the true
(or canonical) symmetries in a unique way. 

We have applied our methods to model building. In the context of GUTs, we have
identified an obstacle to completely forbidding dimension five proton decay
operators in certain \SO{10} GUTs. In  string model building our methods
allow us to identify novel vacuum configurations with an exact $R$ parity. 

\subsection*{Acknowledgements}

We would like to thank R.~Kappl and P.~Vaudrevange for discussions.
This research is supported by the DFG cluster of excellence Origin
and Structure of the Universe, the Graduiertenkolleg ``Particle Physics at the
Energy Frontier of New Phenomena'' and the SFB-Transregios 27 ``Neutrinos and
Beyond'' by the Deutsche Forschungsgemeinschaft (DFG). Some parts of this work
were carried out at the Aspen Center for Physics, which we would like to thank
for the hospitality.


\begin{thebibliography}{10}

\bibitem{Farrar:1978xj}
G.~R. Farrar and P.~Fayet, Phys.\ Lett.\ \textbf{B76} (1978), 575--579.

\bibitem{Dimopoulos:1981dw}
S.~Dimopoulos, S.~Raby, and F.~Wilczek, Phys.\ Lett.\ \textbf{B112} (1982), 133.

\bibitem{Mohapatra:1998rq}
R.~N. Mohapatra and P.~B. Pal, World Sci.\ Lect.\ Notes Phys.\ \textbf{60} (1998),
  1--397.

\bibitem{Lebedev:2007hv}
O.~Lebedev et al., Phys. Rev. \textbf{D77} (2007), 046013,
  [arXiv:0708.2691 [hep-th]].

\bibitem{Krauss:1988zc}
L.~M. Krauss and F.~Wilczek, Phys.\ Rev.\ Lett.\ \textbf{62} (1989), 1221.

\bibitem{Banks:1989ag}
T.~Banks, Nucl.\ Phys.\ \textbf{B323} (1989), 90.

\bibitem{Preskill:1990bm}
J.~Preskill and L.~M. Krauss, Nucl.\ Phys.\ \textbf{B341} (1990), 50--100.

\bibitem{Adulpravitchai:2009kd}
A.~Adulpravitchai, A.~Blum, and M.~Lindner,  (2009),  0907.2332.

\bibitem{Jacobsen}
N.~Jacobson, Basic Algebra I, W. H. Freeman and Company (1974), 471 p.

\bibitem{Dreiner:2005rd}
H.~K. Dreiner, C.~Luhn, and M.~Thormeier, Phys.\ Rev.\ \textbf{D73} (2006),
  075007,  [hep-ph/0512163].

\bibitem{ZNcode:2009zn}
R.~Schieren, \emph{{DiscreteBreaking}},\\
{\texttt{http://einrichtungen.physik.tu-muenchen.de/T30e/codes/DiscreteBreaking/}}.

\bibitem{Mohapatra:2007vd}
R.~N. Mohapatra and M.~Ratz, Phys.\ Rev.\ \textbf{D76} (2007), 095003,
  [0707.4070].

\bibitem{Csaki:1997aw}
C.~Cs{\'{a}}ki and H.~Murayama, Nucl.\ Phys.\ \textbf{B515} (1998), 114--162,
  [hep-th/9710105].

\bibitem{Ibanez:1991hv}
L.~E. Ib{\'a}{\~n}ez and G.~G. Ross, Phys.\ Lett.\ \textbf{B260} (1991),
  291--295.

\bibitem{Banks:1991xj}
T.~Banks and M.~Dine, Phys.\ Rev.\ \textbf{D45} (1992), 1424--1427,
  [hep-th/9109045].

\bibitem{Araki:2008ek}
  T.~Araki et al.,
  Nucl.\ Phys.\   \textbf{B805} (2008) 124
  [arXiv:0805.0207 [hep-th]].

\bibitem{Azatov:2008vu}
A.~T. Azatov and R.~N. Mohapatra, Phys.\ Rev.\ \textbf{D78} (2008), 015002,
  [0802.3906].

\bibitem{Lebedev:2006kn}
O.~Lebedev et al., Phys.\ Lett.\ \textbf{B645} (2007), 88,
  [hep-th/0611095].

\bibitem{Buchmuller:2008uq}
W.~Buchm{\"u}ller and J.~Schmidt, Nucl.\ Phys.\ \textbf{B807} (2009), 265--289,
  [0807.1046].

\bibitem{Buchmuller:2005jr}
W.~Buchm{\"u}ller, K.~Hamaguchi, O.~Lebedev, and M.~Ratz, Phys.\ Rev.\ Lett.\ 
  \textbf{96} (2006), 121602,  [hep-ph/0511035].

\bibitem{Buchmuller:2006ik}
W.~Buchm{\"u}ller, K.~Hamaguchi, O.~Lebedev, and M.~Ratz, Nucl.\ Phys.\ 
  \textbf{B785} (2007), 149--209,  [hep-th/0606187].

\end{thebibliography}
\providecommand{\bysame}{\leavevmode\hbox to3em{\hrulefill}\thinspace}

\end{document}